\documentclass[twocolumn,aps,prl,groupedaddress,footinbib,amsmath,amssymb,showpacs]{revtex4-1}
\usepackage{graphicx}
\usepackage{bm}
\usepackage[colorlinks,	linkcolor=blue,anchorcolor=blue,citecolor=blue,bookmarksnumbered]{hyperref}
\usepackage{float}

\usepackage{amsfonts}
\usepackage{bm}
\usepackage{xcolor}

\begin{document}

\title{Fast spin squeezing by distance-selective long-range interactions
with Rydberg molecule dressing}

\author{Huaizhi Wu}

\affiliation{Fujian Key Laboratory of Quantum Information and Quantum Optics and
Department of Physics, Fuzhou University, Fuzhou 350116, People's
Republic of China}

\author{Xin-Yu Lin}

\affiliation{Fujian Key Laboratory of Quantum Information and Quantum Optics and
Department of Physics, Fuzhou University, Fuzhou 350116, People's
Republic of China}

\author{Zong-Xing Ding}

\affiliation{Fujian Key Laboratory of Quantum Information and Quantum Optics and
Department of Physics, Fuzhou University, Fuzhou 350116, People's
Republic of China}

\author{Shi-Biao Zheng,}

\affiliation{Fujian Key Laboratory of Quantum Information and Quantum Optics and
Department of Physics, Fuzhou University, Fuzhou 350116, People's
Republic of China}

\author{Igor Lesanovsky}

\affiliation{School of Physics and Astronomy, University of Nottingham, Nottingham
NG7 2RD, United Kingdom }

\affiliation{Centre for the Mathematics and Theoretical Physics of Quantum Non-equilibrium
Systems, University of Nottingham, Nottingham NG7 2RD, United Kingdom}

\author{Weibin Li}

\affiliation{School of Physics and Astronomy, University of Nottingham, Nottingham
NG7 2RD, United Kingdom }

\affiliation{Centre for the Mathematics and Theoretical Physics of Quantum Non-equilibrium
Systems, University of Nottingham, Nottingham NG7 2RD, United Kingdom}
\begin{abstract}
We propose a Rydberg molecule dressing scheme to create strong and long-ranged
interactions at selective distances. This is achieved through laser
coupling ground-state atoms off-resonantly to an attractive molecular
curve of two interacting Rydberg atoms. Although dephasing due to Rydberg state
decay occurs in all dressing schemes, an advantage of the molecule
dressing is that a large ratio of dressed interaction to dephasing rate
can be realized at large atomic separations. In an optical lattice or tweezer setting, we show that
the strong interaction permits the fast generation of spin squeezing for several tens of
dressed atoms. The proposed setting offers a new route to study complex
many-body dynamics and to realize quantum information processing with
non-convex long-range interactions. 
\end{abstract}
\maketitle
\textbf{\textit{Introduction.}}\textendash The long lifetime and strong interaction emerging between atoms excited to Rydberg states offer an effective
way for the implementation of quantum computation protocols~\cite{Jaksch2000,Lukin2001,Wilk2010,Zhang2010,Isenhower2010,Keating2013,Hankin2014,Keating2015,Li2016,Levine2018}, and many-body
quantum simulation \cite{Honer2010,Mattioli2013,Glaetzle2014,Glaetzle2015,Schauss1455,Leseleuc2017,Bijnen2015,Li2015,Labuhn2016,Leseleuc2018a}. 
One approach to utilize the Rydberg interaction is through the  Rydberg dressing \cite{Santos2000,Pupillo2010,Henkel2010,Johnson2010,Balewski_2014,Jau2015,Zeiher2016,Mukherjee2016,Buchmann2017,Plodziefmmodecutenlseni2017,Chai2017,PhysRevLett.124.063601},
where ground state atoms are off-resonantly coupled to a Rydberg state. As direct Rydberg excitation is avoided, the dressed interaction - although weakened in strength - comes in principle with the benefit of prolonged coherence times. For Rydberg atoms interacting with a van der Waals potential, the dressing leads to a soft-core interaction (with characteristic soft-core radius $R_c$) between the dressed atoms. This interaction is nearly a constant when the distance $R$ between the atoms is much smaller than  $R_c$ and it decays rapidly when $R>R_c$. The strength and $R_c$ of the dressed interaction can be controlled by the Rydberg dressing laser (i.e. detuning, Rabi frequency and the coupled Rydberg state).   
Such tunability~\cite{Carroll2004,Barredo2014} has stimulated the study of
exotic topological states~\cite{Glaetzle2014,Glaetzle2015,Bijnen2015,Shi2018}, density-wave orders~\cite{Khasseh2017,Li2018,PhysRevLett.124.140401}, correlated spin dynamics~\cite{Zeiher2017,PhysRevX.11.021036}, and quantum sensing \cite{Bouchoule2002,Gil2014,PhysRevLett.122.053601} with Rydberg dressed interactions.

\begin{figure}
\includegraphics[width=1\columnwidth]{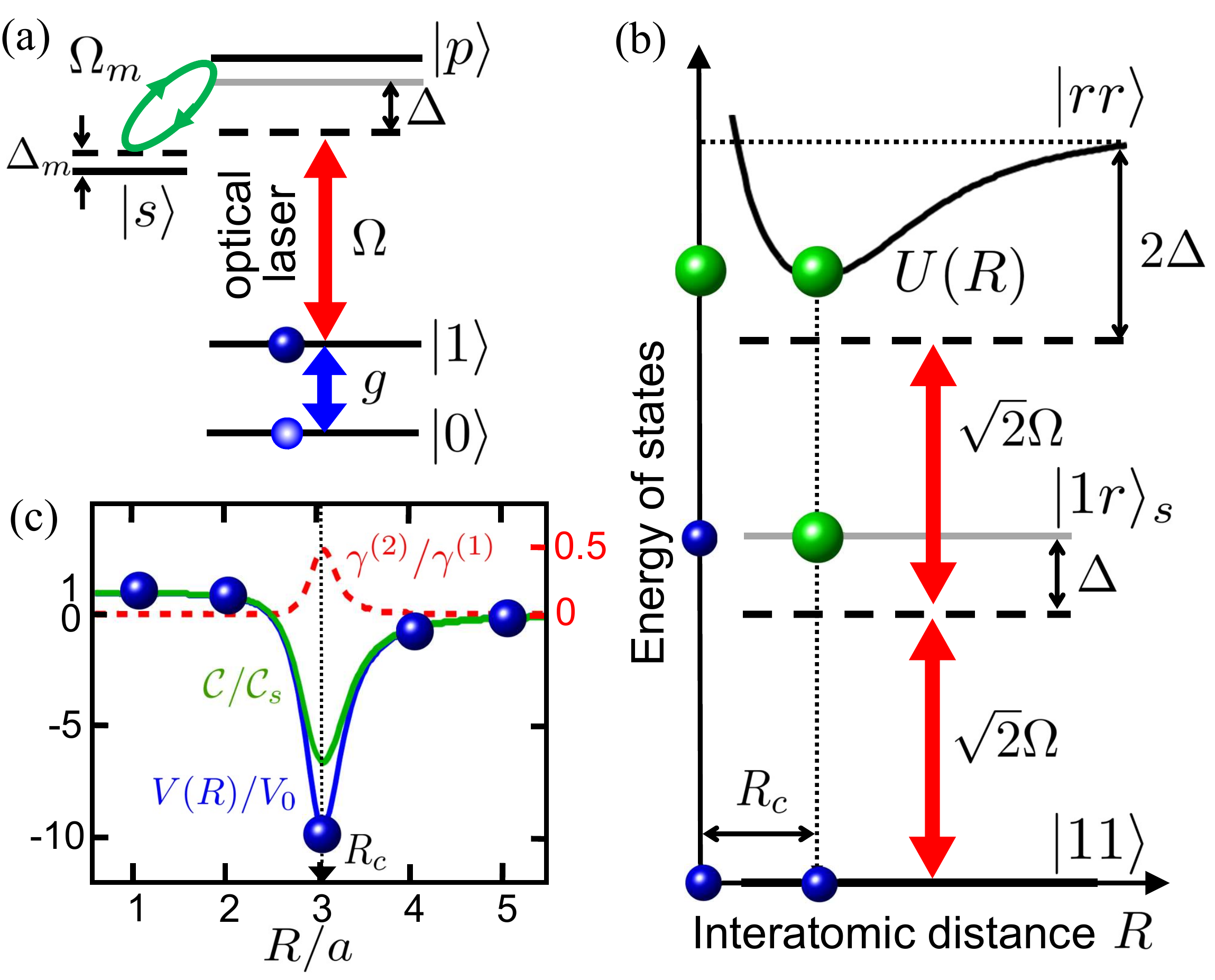}
\caption{\label{fig:Model_Scheme} \textbf{Rydberg molecule dressing}.
(a) Level scheme. The ground states $|0\rangle$ and $|1\rangle$
 are coupled via optically Raman transitions.  The ground state $|1\rangle$ is optically dressed
to a Rydberg state $|r\rangle$. Rydberg state $|nS\rangle$ and $|n'P\rangle$ are coupled by a microwave field (Rabi frequency $\Omega$
and detuning $\Delta$), forming a superposition state $|r\rangle$. (b) Rydberg molecule dressing. Two-photon
transition from the ground state $|11\rangle$ to the Rydberg molecular
dimer state $|rr\rangle$ is detuned by $2\Delta+U(R)$, with $U(R)$
being the attractive interaction potential between atoms in Rydberg state $|r\rangle$. (c) RMD
induced interaction $V(R)$, nonlocal two-body decay $\gamma^{(2)}$,
and coherence strength $\mathcal{C}$ as a function of the interatomic
distance $R$. 
For comparison, $V(R)$, $\mathcal{C}$, and $\gamma^{(2)}$ are shown in units
of $V_{0}$,  $\gamma^{(1)}$ and $\mathcal{C}_s$, respectively. Parameters are $(\Omega,\Delta,\gamma)/2\pi=(1,10,0.01)$
MHz, $\Delta_2(R_c)=2\Omega$, and lattice spacing $a=1\,\mu$m.  See text for details.}
\end{figure}

Dissipation~\cite{Weber2012,Malossi2014,Schempp2014,Goldschmidt2016,Boulier2017},
such as spontaneous decay due to the finite lifetime of Rydberg states~\cite{Barredo_2020}
and motional dephasing~\cite{li_nonadiabatic_2013,Gambetta_2020,lorenz_raman_2021}, is ubiquitous in any Rydberg dressing scheme. In the weak
Rydberg dressing scheme (i.e. $|\Delta|\gg\Omega$ with $\Delta$ and $\Omega$ the detuning and Rabi frequency of the dressing laser)~\cite{Henkel2010,Pupillo2010,Keating2013,Zeiher2016}, the effective decay rate of the dressed state scales with $\sim\Omega^{2}\gamma/\Delta^{2}$ with $\gamma$ to be the spontaneous decay rate in Rydberg states. As a result, the effective lifetime of the dressed atom is increased by a factor $\Delta^{2}/\Omega^{2}$.
The strength of the dressing interaction scales as $V_0\sim\Omega^{4}\Delta^{-3}$. Its coherence strength is given by the ratio between
$V_0$ and the effective decay rate, $\mathcal{C}\propto\Omega^{2}/(\Delta\gamma)$~\cite{Keating2016,Zeiher2016,Lee2017a}.
These estimations, however, are based on a simple two-level description. There are other detrimental processes such as level crossings at short distances that have to be taken care of in an experimental implementation~\cite{Zeiher2016}. To effectively implement quantum simulation and computation with the dressed long-range interactions~\cite{Gross995,Schauss_2018,wu_concise_2021}, it is worth to developing
new Rydberg dressing schemes where such unwanted effects can be suppressed, and which offers a fundamentally better scaling of the  coherence strength $\mathcal{C}$. 

In this work, we propose a Rydberg dressing scheme
where ground state atoms are coupled off-resonantly to macroscopic
Rydberg molecular dimers which form attractive potentials at micrometer
separations~\cite{petrosyan_binding_2014}. This results in a long-range
dressed interaction that is strongly attractive 
only at selective distances, while saturating to $V_{0}$ at 
short distances ($R\ll R_c$). The selective distance locates at the minimal of
the molecular potential, whose depth and therefore the dressed interaction
can be modulated by external microwave fields. Though laser dressing to Rydberg states generally leads
to single- and two-body dephasing, we show that the Rydberg molecule
dressing (RMD), however, greatly enhances $\mathcal{C}$. As an application, we study spin squeezing of the Rydberg dressed atom in an one dimensional optical lattice. We show that
this allows to implement stronger spin squeezing within a much shorter time 
in comparison to ''traditional" soft-core Rydberg dressing (SRD)~\cite{Henkel2010,Pupillo2010}, provided that the atoms are positioned at certain preselected distances. Our
study reveals insights on the coherent and dissipative features in
Rydberg gases, which are important for implementing many-body physics
and quantum information processing with Rydberg dressed interactions~\cite{Honer2010,Mattioli2013,Glaetzle2014,Glaetzle2015,Bijnen2015,Li2015}.

\textbf{\textit{Rydberg molecule dressing.}}\textendash In our setting,
each atom is modeled by two hyperfine ground states $|0\rangle$
and $|1\rangle$, coupled through a Raman process with strength $g$,
as shown in Fig.~\ref{fig:Model_Scheme}(a). A microwave field with
the Rabi frequency $\Omega_{m}$ and the detuning $\Delta_{m}$ couples
Rydberg $nS$ and $n'P$ states (where $n$ and $n'$ are principal
quantum numbers) ~\cite{Arias2019}. This leads to a Rydberg superposition
state $|r\rangle$. The dispersion coefficients of the two Rydberg
states are assumed to have opposite signs. In this case, two atoms in the superposition state $|r\rangle$ experience an interaction potential
$U(R)$ with an attractive tail and a repulsive core, depicted in Fig. \ref{fig:Model_Scheme}(b). More details can be found in the supplementary material (\textbf{SM}) \cite{SM}. The minimum (located at $R_{c}$) of $U(R)$
can be a few micrometers, supporting exotic many-body states
and macroscopic dimers~\cite{petrosyan_binding_2014}.

In the proposed RMD scheme, the hyperfine state $|1\rangle$ is coupled to state $|r\rangle$ where the laser
is red-detuned with respect to the potential minimum, as illustrated in Fig.~\ref{fig:Model_Scheme}(b)$  $. When $\Delta\gg\Omega$ and the two-photon
detuning $\Delta_{2}(R)=U(R)+2\Delta$ is much larger than the effective
Rabi frequency $\Omega_{2}\equiv\left(\sqrt{2}\Omega\right)^{2}/2\Delta$
of the collective transition $|11\rangle\leftrightarrow|rr\rangle$
(via the intermediate state $|1r\rangle_{s}\equiv(|1r\rangle+|r1\rangle)/\sqrt{2}$),
i.e. $|\Delta_{2}(R_{c})|\gg\Omega_{2}$, we obtain an effective interaction
potential between two atoms in the dressed $|1\rangle$ state, 
\begin{eqnarray}
V(R)=\frac{V_0\Delta_{2}(R)}{\Delta_{2}^{2}(R)+\gamma^{2}}U(R),\label{eq:dressed_potential-2}
\end{eqnarray}
with $V_{0}\equiv\Omega^{4}/8\Delta{}^{3}$ and $\gamma$ being the spontaneous decay rate in state $|r\rangle$~\cite{SM}. As $|\Delta_{2}(R)|\gg\gamma$, the dressed potential
can be further approximated to be $V(R)\approx V_{0}U(R)/\Delta_{2}(R)$.
The dressed potential is repulsive and attractive 
at short and large distances [see Fig.~\ref{fig:Model_Scheme}(c)], respectively.
Its depth is largest around the distance $R_{c}$ determined by $U(R_{c})/\Delta_{2}(R_{c})$.
Note that the interaction strength at $R_{c}$ can be several times
larger than the saturated value ($\sim V_{0}$) at short distances.
Such profile is different from the soft-core potential~\cite{Jau2015,Keating2016,Zeiher2016,Lee2017a}.

The Rydberg dressing induces not only the dispersive interaction, but also dissipation, whose origin lies in the
finite lifetimes in Rydberg states. In our system, the Rydberg dressing laser
induces a single-body dephasing (SBD) with the rate $\gamma^{(1)}=\Omega^{2}\gamma/4\Delta^{2}$
and a distance dependent two-body dephasing (TBD) with a rate given
by 
\begin{eqnarray}
\gamma^{(2)}(R)=\frac{\Omega^{4}\gamma}{2\Delta^{2}[\Delta_{2}^{2}(R)+\gamma^{2}]}.
\end{eqnarray}
This two-body dephasing rate reaches the maximal value at $R=R_{c}$.

Dephasing is ubiquitous in any Rydberg dressing scheme. However, the advantage of the RMD is that the interaction strength $V(R)$ can
be enhanced by an order of magnitude with the dephasing only being
increased by about a factor of two. To illustrate this, we define
the interaction coherence strength $\mathcal{C}$ 
\begin{equation}
\mathcal{C}\equiv\frac{V(R)}{2\gamma^{(1)}+\gamma^{(2)}(R)}\approx\frac{\Omega^{2}}{4\Delta\gamma}\cdot\frac{\Delta_{2}(R)U(R)}{\Omega^{2}+\Delta_{2}^{2}(R)},\label{eq:ratio_E2Gamma1}
\end{equation}
where we have taken into account the SBD and TBD. The value of $\mathcal{C}$
strongly depends on the laser parameters. When $\left|\Delta_{2}(R)\right|\sim\left|\Delta\right|\gg\Omega$,
one finds the familiar scaling for
the SRD $\mathcal{C}_{s}=V_0/2\gamma^{(1)}=\Omega^{2}/4|\Delta|\gamma$. For $\left|\Delta\right|\gg\left|\Delta_{2}(R)\right|\sim\Omega$,
the ratio becomes $|\mathcal{C}|=\left|\Delta_{2}(R)\right|/4\gamma$, which
can be large. For example, we can choose $|\Delta_{2}(R)|=\Omega$, such
that $\mathcal{C}_{m}=\Omega/4\gamma\sim|\Delta|/\Omega\mathcal{C}_{s}$. As typically
$|\Delta|\gg\Omega$, one can gain an order of magnitude increasing
in $\mathcal{C}$ at
$R=R_{c}$, depicted in Fig. \ref{fig:Model_Scheme}(c).

\begin{figure}
\includegraphics[width=1\columnwidth]{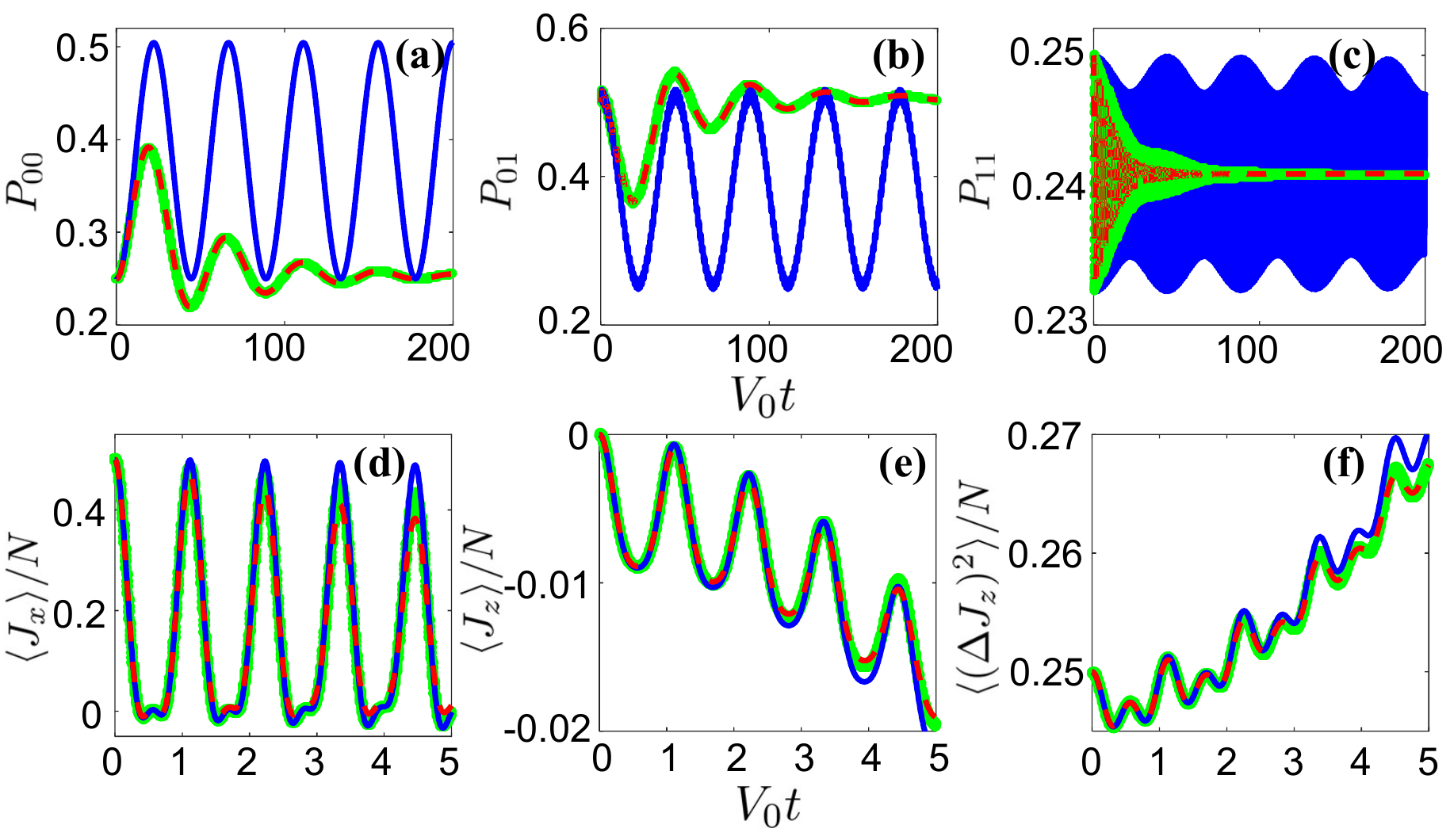}
\caption{\label{fig:Dyn_Eff_ME} \textbf{Driven-dissipative dynamics}.
(a)-(c) Population of the two atom states $|0_{1}\rangle|0_{2}\rangle$, $(|0_{1}\rangle|1_{2}\rangle+|1_{1}\rangle|0_{2}\rangle)/\sqrt{2}$,
and $|1_{1}\rangle|1_{2}\rangle$ versus the scaled time $V_{0}t$
for $N=2$. (d)-(f) Time-dependence
of the mean spin components $\langle J_{x}\rangle$, $\langle J_{z}\rangle$,
and the covariance $\langle(\Delta J_{z})^{2}\rangle$ for $N=7$.
Atoms are initially in the state $\otimes_{k=1}^{N}(|0_{k}\rangle+|1_{k}\rangle)/\sqrt{2}$.
We show numerical data with (red dash) and without (green solid) TBD,
as well as the coherent dynamics (blue solid) by turning off both the SBD and TBD. Other parameters are
$\Omega=1$, $(\Delta,\gamma)/\Omega=(5.5,0.005)$, $R_{c}=a$, $g/V_{0}=0.2$,
$n=50$, and $n^{\prime}=58$. }
\end{figure}

\textbf{\textit{Effective many-body spin model of dressed Rydberg atoms.}}\textendash We now consider
the many-body dynamics of  the atoms trapped in a deep one-dimensional (1D)
optical lattice~\cite{zhang_magic-wavelength_2011,li_entanglement_2013}. The dynamics with the dressed ground state manifold is then described by Hamiltonian
$\mathcal{H}_{s}=\sum_{k=1}^{N}[gJ_{x}^{(k)}+\delta^{(k)}J_{z}^{(k)}]+\sum_{k<l}V(R_{kl})J_{z}^{(k)}J_{z}^{(l)}$
with the spin-$1/2$ Pauli operators $J_{x}^{(k)}=\left(|1_{k}\rangle\langle0_{k}|+|0_{k}\rangle\langle1_{k}|\right)/2$,
$J_{y}^{(k)}=i\left(|0_{k}\rangle\langle1_{k}|-|1_{k}\rangle\langle0_{k}|\right)/2$,
and $J_{z}^{(k)}=\left(|1_{k}\rangle\langle1_{k}|-|0_{k}\rangle\langle0_{k}|\right)/2$.
The dressing-induced interaction between atoms located on lattice
sites $k$ and $l$ at a distance $R_{kl}/a=|k-l|$ is given by $V(R_{jk})$.
Taking into account of the single and two-body dephasing, the dynamics
of the  many-body two-level system is governed by the master
equation, 
\begin{equation}
\stackrel{.}{\rho}=-i[\mathcal{H}_{s},\rho]+\sum_{k=1}^{N}\mathcal{L}[o_{k}]\rho+\sum_{k<l}\mathcal{L}[o_{kl}]\rho,\label{eq:ME_eff_deph}
\end{equation}
where the Lindblad operator $\mathcal{L}[o]\rho=o\rho o^{\dagger}-\frac{1}{2}\{o^{\dagger}o,\rho\}$ with 
$o_{k}=\sqrt{\gamma^{(1)}}|1_{k}\rangle\langle1_{k}|$, and  $o_{kl}=\sqrt{\gamma^{(2)}(R_{kl})}|1_{k}1_{l}\rangle\langle1_{k}1_{l}|$ being the so-called jump operator corresponding to single- and two-body dephasing, respectively. 

We first explore the competition between the dephasing and dressed two-body interaction with an example of two atoms, shown in Fig. \ref{fig:Dyn_Eff_ME}(a)-(c). During the early stage of dynamical evolution $0<V_0 t<10$,  both the SBD and TBD hardly affect the coherent dynamics. This is seen from the population dynamics, where the  master equation simulation agrees with the coherent dynamics.  They only become important in the long-time dynamics, where the many-body coherence is damped. In this region, the dominant contribution results from the SBD, as found from our numerical simulation. 

The short-time dynamics is largely coherent even for a moderate number of spins. In Fig.
\ref{fig:Dyn_Eff_ME}, the expectation values of the collective spin
operator $J_{x}=\sum_{k}J_{x}^{(k)}$ and $J_{z}=\sum_{k}J_{z}^{(k)}$,
and the quantum fluctuation $\langle(\Delta J_{z})^{2}\rangle$ of
the operator $J_{z}$ for a chain of $N=7$ atoms are shown. The numerical simulations show that the dephasing  plays visible roles when $V_0t>4$.  When $V_0t<4$,  the high coherence strength guarantees that the dispersive interaction dominates in
the short-time dynamics.

\textbf{\textit{Fast spin squeezing.}}\textendash We will show that the large $\mathcal{C}$ in the RMD 
allows to achieve rapid spin squeezing in the 1D lattice~\cite{Ma2011}.
The degree of spin squeezing can be quantified
through the squeezing parameter (SP)~\cite{Wineland1994}, 
\begin{equation}
\xi^{2}=\frac{N\left(\Delta J_{\vec{n}_{\bot}}\right)^{2}}{|\langle\vec{J}\rangle|^{2}},\label{eq:Squeezing_Wineland}
\end{equation}
where $J_{\vec{n}_{\bot}}$ is the spin component perpendicular to
the direction of the mean total spin $\langle\vec{J}\rangle$ with
$\vec{J}=\sum_{k}\boldsymbol{J}{}^{(k)}$. The parameter $\xi^{2}$
is then defined by choosing the direction $\vec{n}_{\bot}$ at which
$\Delta J_{n_{\bot}}$ is minimized. For the state with $\xi^{2}<1$,
its phase sensitivity is improved over the standard quantum limit
($\sim1/\sqrt{N}$) given by the coherent spin state~\cite{qin_strong_2020}.

To achieve spin squeezing, initially the atoms are prepared in the
state $|0\rangle$ such that the mean spin direction is along the
$z$-direction, and thus the perpendicular spin component is simply
$J_{\vec{n}_{\bot}}\left(\theta\right)=\textrm{cos}\left(\theta\right)J_{x}+\textrm{sin}\left(\theta\right)J_{y}$
with $\theta$ being the angle with respect to the $x$ axis, which
immediately implies $\left(\Delta J_{\vec{n}_{\bot}}\right)^{2}=\textrm{cos}^{2}\left(\theta\right)\langle J_{x}^{2}\rangle+\textrm{si\ensuremath{n^{2}}}\left(\theta\right)\langle J_{y}^{2}\rangle+\text{sin}\left(2\theta\right)\langle J_{x}J_{y}+J_{y}J_{x}\rangle/2$.
We then implement the spin-echo type squeezing protocol \cite{Gil2014},
which consists of a  $\pi/2$ pulse of length $t_{\pi/2}=\pi/\left(2g\right)$
to rotate the spin along the $x$ axis,  RMD
of duration $\tau/2$, a $\pi$ pulse to rotate the spin, 
RMD of duration $\tau/2$, and then a $\pi/2$ pulse.
During the spin rotation, the Rydberg dressing laser is turned
off. At the end of the sequence, the mean spin is along the $z$-direction
again with the noise uncertainty for $J_{\vec{n}_{\bot}}$ along one
of the perpendicular direction being squeezed.

The SP resulting from the above protocol can be analytical calculated in the absence of dissipation~\cite{Gil2014}. In the presence of dissipation, it is not possible to carry out analytical calculations
 for the dressed spin model. For small systems with a few atoms we can study the driven-dissipative
dynamics by numerically solving the master equation and evaluate the
SP. Approximations have to be taken when there are tens and even hundreds of lattice sites due to the large Hilbert space.
Since the $\pi/2$ and $\pi$ rotational pulses are much shorter than the typical
time scale of the SBD and TBD, dissipation is therefore not important during the dynamical evolution. This allows us to simulate the dynamics approximately with
a non-Hermitian Hamiltonian $\mathcal{H}_{c}=\mathcal{H}_{s}-i(\gamma^{(1)}/2)\sum_{k=1}^{N}|1_{k}\rangle\langle1_{k}|-\left(i/2\right)\sum_{j<k}\gamma^{(2)}(R_{jk})|1_{j}1_{k}\rangle\langle1_{j}1_{k}|$, where the single- and two-body jump processes have been neglected. Furthermore, we apply the mean-field approximation
to the decoherence part induced by the TBD, and finally arrive at the non-Hermitian Hamiltonian
\begin{eqnarray}
\mathcal{H}_{c} & \approx & \frac{1}{2}\sum_{j=1}^{N}\left(\sum_{k\neq j}V(R_{jk})-i\Gamma_{z}\right)J_{z}^{(j)}\nonumber \\
 &  & +\sum_{j<k}V(R_{jk})J_{z}^{(j)}J_{z}^{(k)}-i\frac{\bar{\Gamma}}{2},\label{eq:H_c}
\end{eqnarray}
where the mean-field decay rate $\bar{\Gamma}=\frac{N}{2}[\gamma^{(1)}+\Gamma_0(\frac{1}{4}-\langle {J}_{z}\rangle{}^{2}/N^2)]$
and $\Gamma_{z}=\gamma^{(1)}+{\Gamma}_0(\frac{1}{2}+\langle {J}_{z}\rangle/N)$  with
$\Gamma_0=\sum_{|k-j|=1}^{N-1}\gamma^{(2)}(R_{jk})$.  We then evaluate the time-dependent many-body state $|\psi(t)\rangle\approx e^{-i\mathcal{H}_c t}|\psi(0)\rangle$ where the initial state $|\psi(0)\rangle$ corresponds to all atoms in state $|0\rangle$. 
\begin{figure}
	\includegraphics[width=1\columnwidth]{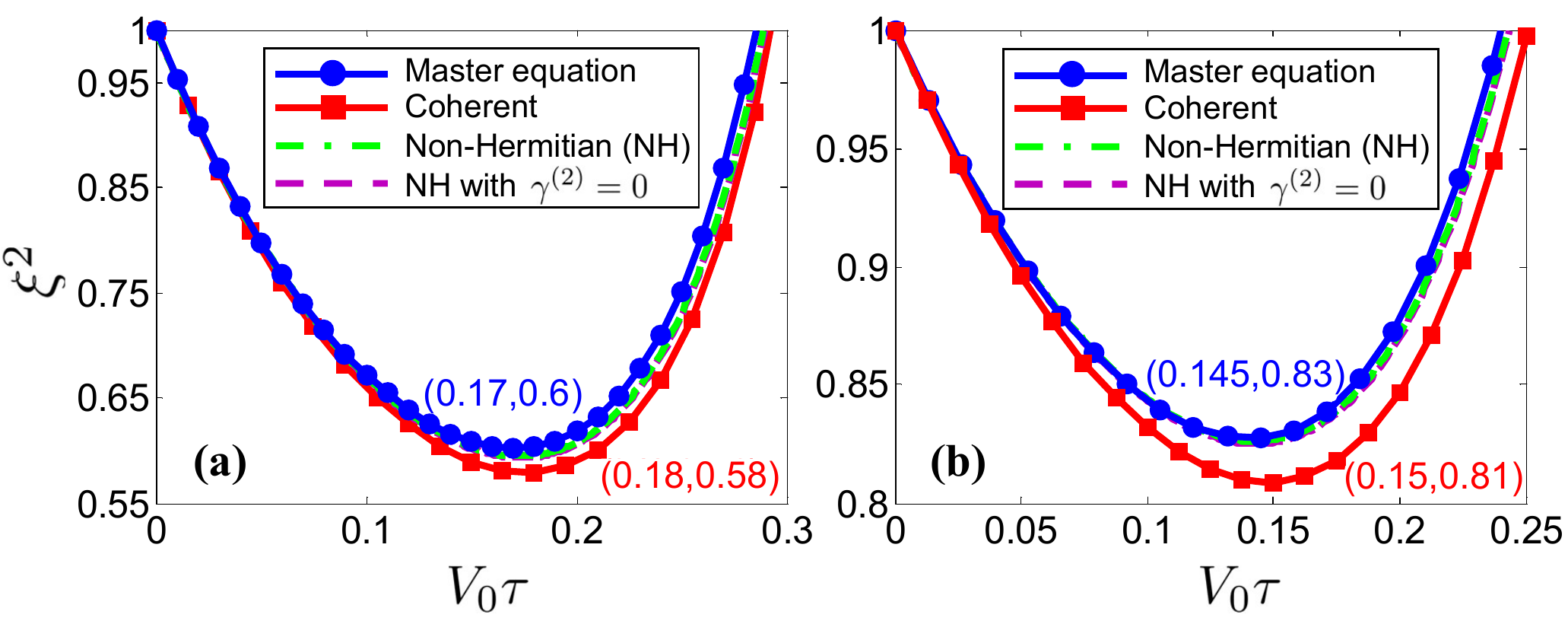}\caption{\label{fig: few_spin_squeezing}\textbf{The squeezing parameter as a function of the dressing time}.  Numerical results obtained
		from the master equation (\ref{eq:ME_eff_deph}), the conditional
		Hamiltonian with and without the two-body decay, and the coherent
		dynamics, are shown. We consider two different lattice constants (a)
		$R_{c}=a$ and (b) $R_{c}=3a$. The coordinates for the optimal coherent
		and dissipative squeezing (i.e. the minimum) are labeled. Here $N=10$ and other parameters
		are the same as in Fig. \ref{fig:Dyn_Eff_ME}. }
\end{figure}

\begin{figure}
\includegraphics[width=1\columnwidth]{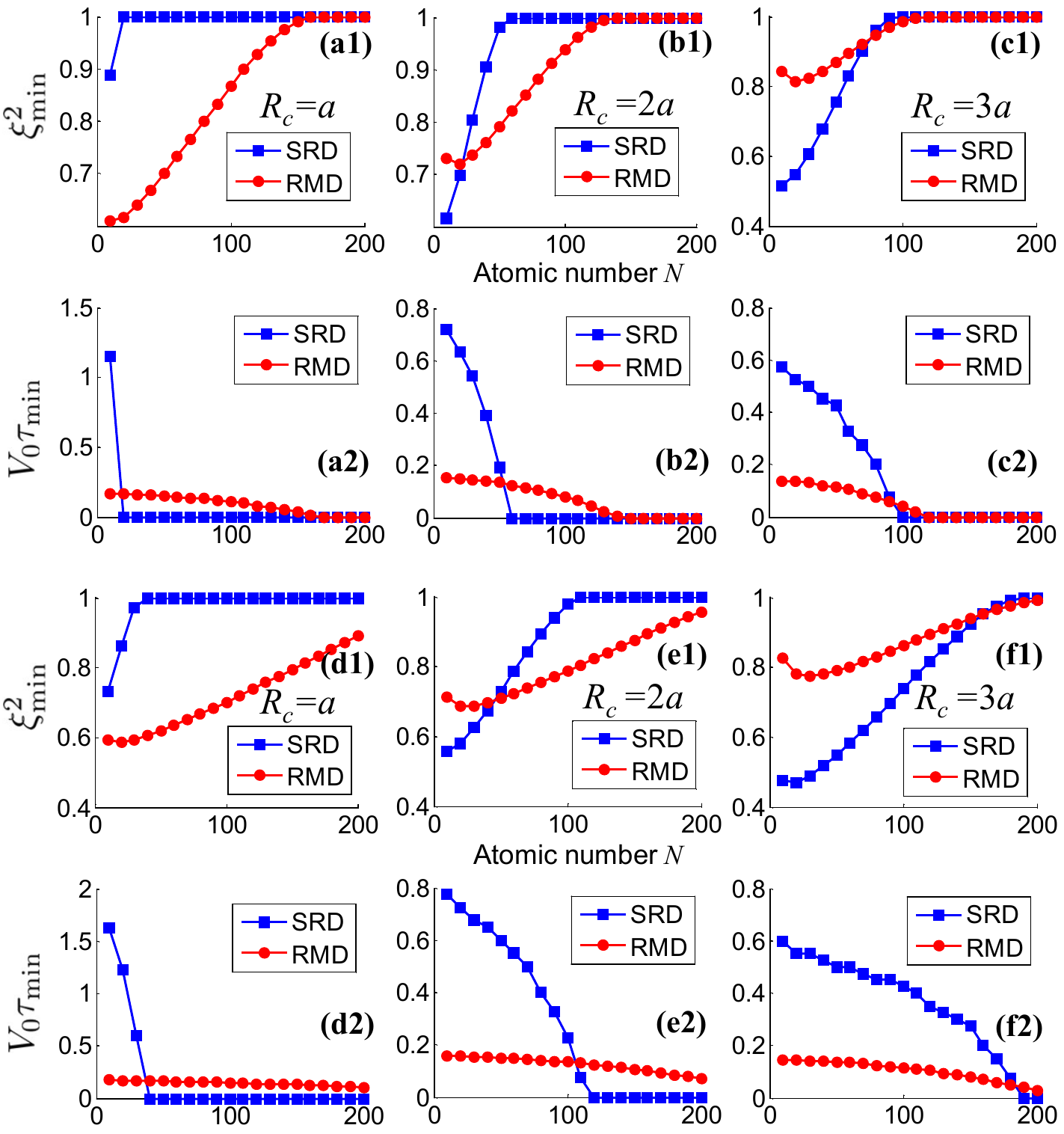}
\caption{\label{fig: Scaling_Na}\textbf{Spin squeezing with RMD and SRD}. The optimal spin squeezing $\xi_{\text{min}}^{2}$
{[}(a1)-(f1){]} and the required dressing time $V_{0}\tau_{\text{min}}$
{[}(a2)-(f2){]} of the RMD and SRD schemes are shown. For $V_{0}\tau_{\text{min}}=0$
in (a2)-(f2), the squeezing parameter $\xi^{2}$ remains larger than
1 for all interaction time $\tau$. The atomic decay rates are $\gamma/\Omega=\text{ }0.01$
in (a)-(c), and $\gamma/\Omega=\text{ }0.005$ in (d)-(f). Other
parameter are the same as in Fig. \ref{fig: few_spin_squeezing}.}
\end{figure}

Eq. (\ref{eq:H_c}) enables us to find an analytical solution
in the presence of the dephasing and for large systems [see \textbf{SM}~\cite{SM}].
To verify the accuracy of the analytical solution against the exact result, we first examine a situation with $N=10$ atoms with $(\Delta,\gamma)/\Omega=(10,0.005)$
(corresponding to $\gamma^{(1)}/\gamma\approx8.3\times10^{-3}$), depicted in Fig.~\ref{fig: few_spin_squeezing}.
For $R_{c}=a$, both the effective master equation~\cite{CIVITARESE2009754,Kominis2008}
and the non-Hermitian Hamiltonian give the similarly optimal SP (i.e.
the minimum) $\xi_{\text{min}}^{2}\sim0.6$ around the dressing time
$V_{0}\tau_{\text{min}}\approx0.17$, which is much shorter than that
by the traditional SRD scheme~\cite{Gil2014}, and see Fig.~\ref{fig: Scaling_Na}(a). The dissipative SP is
slightly weaker than the coherent counterpart $\xi^{2}\sim0.58$ without
including the decoherence. For $R_{c}=3a$, because the repulsive
interactions among the nearest and the second-nearest neighboring
sites counteract the effect of the strong attractive interactions,
the optimal SP reduces to $\xi_{\text{min}}^{2}\sim0.83$. On the
the hand, in comparison with the individually spontaneous decay, the
TBD does not cause significant effects on the rapid spin squeezing
at the time scale $V_{0}\tau_{\text{min}}$. We find that the
dissipative dynamics is well captured by the conditional Hamiltonian
$\mathcal{H}_{c}$ in general, which motivates us to investigate
larger systems with $\mathcal{H}_{c}$ (see
\textbf{SM}~\cite{SM}).


Based on Eq.~(\ref{eq:H_c}), it is not difficult to find that
the time-dependent squeeze parameter $\xi^{2}(\tau)$ exhibits an
exponential decay as the dressing time $\tau$ increases, namely,
$\xi^{2}(\tau)\propto e^{-\bar{\Gamma}\tau}$, with $\bar{\Gamma}$ depending
on the effective single- and two-body dephasing rates. We 
show in Fig. \ref{fig: Scaling_Na} the optimally dissipative SP $\xi_{\text{min}}^{2}$
and the corresponding dressing time $V_{0}\tau_{\text{min}}$ versus
the number of the lattice sites 
with different lattice constant $R_{c}=ka$ ($k=1,2,3$). For comparison,
we also show the SP influenced only by the SRD, where the TBD is negligible \cite{Gil2014}. In
both cases, the detrimental effects of the dephasing are to diminish  
the squeezing when the number of atoms increases. In particular, the spin squeezing is more robust in case of RMD, where the SP is smaller than 1 for $N$ up to 200 [see the first and third row in Fig.~\ref{fig: Scaling_Na}]. In contrast the SP reaches 1 typically at smaller $N$ with the SRD. On the other hand, for the RMD scheme with
$R_{c}=a$, where the dressed atoms next to each other get the strongest
interaction coherence strength $\mathcal{C}$, the optimal SP exists for $N>150$
and the dressing times are all within $V_{0}\tau_{\text{min}}<0.17$;
the SRD scheme however allows for $\xi_{\text{min}}^{2}<1$ when the
system contains only a few tens of atoms and requires a much longer
dressing time. As the lattice constant decreases from $a=R_{c}/2$ to $R_{c}/3$, the SP for the RMD scheme decreases. This reduction results from the fact that the phase of the many-body state $|\psi(t)\rangle$ due to
repulsive {[}$V(R)>0${]} and attractive {[}$V(R)<0${]}
interactions cancels. In contrast, the SRD scheme becomes more robust
to the dephasing for $a=R_{c}/3$ due to the constructive
phase accumulation by all repulsive interactions (i.e. the interaction potential is positive for all realized interatomic distances). In general, a compromise could
be found in between getting a large interaction coherence strength and
keeping all negative (or positive) phase terms. In this regard,
the SP under the RMD can be better than that by the SRD
for $N>40$ and $R_{c}=2a$ , but the SRD scheme is able to achieve
 strong SP, i.e., $\xi_{\text{min}}^{2}<0.5$ for $R_{c}\geq3a$ and $N<30$.

\textbf{\textit{Conclusion.}}\textendash We have proposed a Rydberg
molecule dressing scheme, where the dressing laser is tuned off-resonantly
to be the attractive molecular potential of two Rydberg atoms, which
can be created by coupling two different Rydberg states with MW fields.
The Rydberg molecule dressing leads to distance-selective attractive
interaction potential of the Rydberg dressed ground-state atoms. We have
shown that enhanced Rydberg dressed interactions can be achieved,
while the respective dissipation is still weak. This allows us to generate entangling
phases for fast spin squeezing. Our work is relevant
to the study of quantum computation and simulation with finite systems
of Rydberg atoms trapped in optical arrays~\cite{labuhn_tunable_2016,bernien_probing_2017,browaeys_many-body_2020} and trapped Rydberg ions~\cite{zhang_submicrosecond_2020}. 
In the future, it is worth to explore squeezing in higher dimensional
settings and investigate whether the distance selective two-body interaction also here
lead to enhanced squeezing. The controllable interaction and dephasing furthermore
opens opportunities to quantum simulate interesting
many-body dynamics, complementary to other studies that aim at the creation of exotic
phases and many-body states~\cite{Glaetzle2017,Glaetzle2017a,Sandor2017,thomas_experimental_2018}. We note that a recent experiment has demonstrated the distance selection interaction in an atom array~\cite{hollerith_realizing_2021}, where the ground state atom is dressed to Rydberg macrodimers~\cite{boisseau_macrodimers_2002}. Our scheme offers the flexibility to engineer the dressed interaction through tuning the MW field~\cite{petrosyan_binding_2014}. 

\begin{acknowledgments}
\textbf{\textit{Acknowledgments.}}   IL  acknowledges support from the Baden-W\"urttemberg Foundation, 
through the project BWST$\_$ISF2019-23 (``Internationale Spitzenforschung")
and the Deutsche Forschungsgemeinschaft through SPP 1929 GiRyd, Grant No. 428276754.  W. L. acknowledges support from
 the EPSRC Grant No. EP/R04340X/1
via the QuantERA project \textquotedblleft ERyQSenS\textquotedblright, and the
Royal Society through the International Exchanges Cost Share award
No. IEC$\backslash$NSFC$\backslash$181078. H.W. and S.B.Z. acknowledge support from the National Natural
Science Foundation of China under Grants No. 11774058, No.
11874114, and No. 12174058.
\end{acknowledgments}

\bibliography{rydbergphysics2}

\onecolumngrid
\clearpage
\widetext
\begin{center}

\textbf{\large Supplementary material for: ``Fast spin squeezing by distance-selective
long-range interactions with Rydberg molecule dressing''}
\end{center}
\setcounter{equation}{0}
\setcounter{figure}{0}
\setcounter{table}{0}
\setcounter{page}{1}
\makeatletter
\renewcommand{\theequation}{S\arabic{equation}}
\renewcommand{\thefigure}{S\arabic{figure}}
\renewcommand{\bibnumfmt}[1]{[S#1]}
\renewcommand{\citenumfont}[1]{S#1}

\begin{center}
    This supplementary material contains additional details on the analysis
in the main text.
\end{center}

\section{microwave modulated Rydberg interaction energies and Rydberg
molecular dressing}
\begin{figure}
\begin{centering}
\includegraphics[scale=0.85]{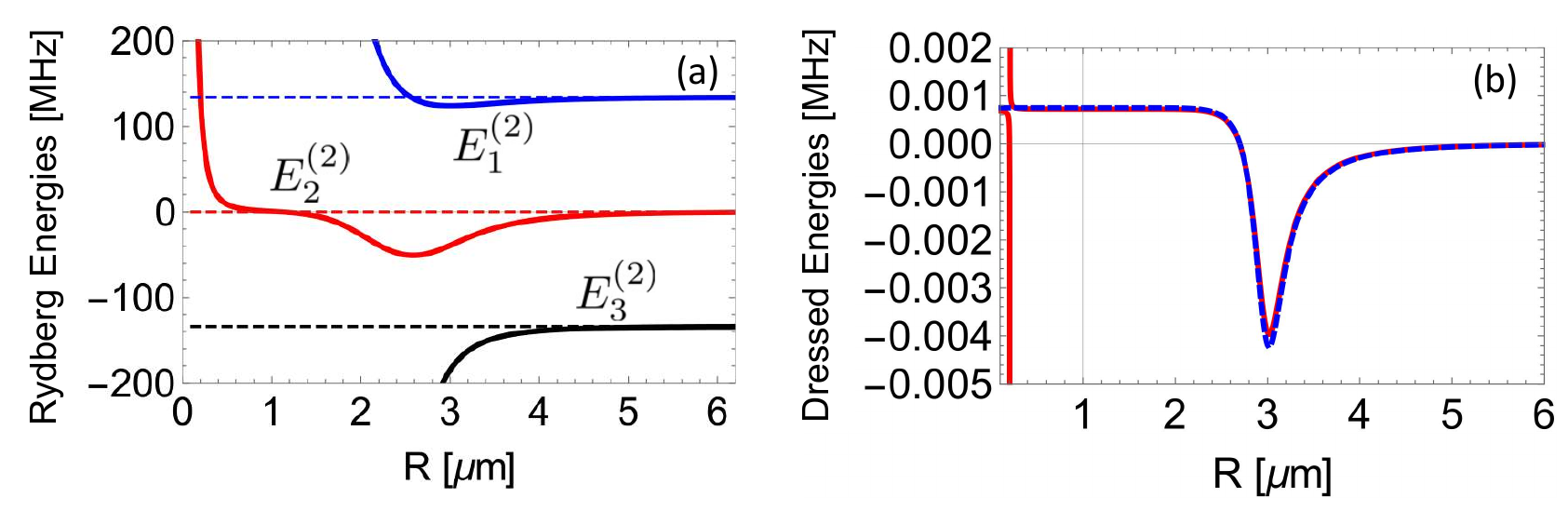} 
\par\end{centering}
\caption{\label{fig:LevelConfig=00003D000026IP}(Color online) (a) Potential energy
$E_{n}^{(2)}(R)$ for the coupled Rydberg pair states $|58p\rangle$
and $|50s\rangle$. The depth of the Rydberg dressed potential can
be modulated by the MW field driving strength and detuning. (b) Interaction
potential for two ground state atoms dressed to the Rydberg state that corresponds to the $E_2^{(2)}$ potential. The dressed interaction potential with full numerical
simulation (red solid) agrees well with the perturbation result
(blue dashed).
The avoided crossing at the very short distance (e.g. smaller than the lattice spacing) is due to the Rydberg
antiblockade, where $|E_{n}^{(2)}(R)-2\Delta_{+}|=0$. Parameters
are $\Omega_{mw}/2\pi=134$ MHz, $\Delta_{mw}=0$. }
\end{figure}

We consider two interacting Rydberg atoms  separated by $R$,
where two Rydberg states $|np\rangle$ and $|n^{\prime}s\rangle$
are continuously driven by a microwave (MW) field with Rabi frequency
$\Omega_{mw}$ and detuning $\Delta_{mw}$. The Rydberg interaction
between the states $|np\rangle$ and $|n^{\prime}s\rangle$ is of
the dipole nature and scales as $R^{-3}$, while the van der Waals
interaction scaling as $R^{-6}$ takes over if the atoms simultaneously
populate the same Rydberg level $|n^{\prime}s\rangle$ (or $|np\rangle)$.
The Hamiltonian for the coupled Rydberg pair states $\{|n's,n's'\rangle, (|n's,np\rangle + |np,n's\rangle)/\sqrt{2}, |np,np\rangle\}$ reads 
\begin{equation}
H_{sp}(R)=\left[\begin{array}{ccc}
-\frac{C_{6}^{(s,s)}}{R^{6}} & \frac{\Omega_{mw}}{\sqrt{2}} & 0\\
\frac{\Omega_{mw}}{\sqrt{2}} & \Delta_{mw}+\frac{C_{3}^{(s,p)}}{R^{3}} & \frac{\Omega_{mw}}{\sqrt{2}}\\
0 & \frac{\Omega_{mw}}{\sqrt{2}} & 2\Delta_{mw}-\frac{C_{6}^{(p,p)}}{R^{6}}
\end{array}\right].\label{eq:H_sp}
\end{equation}
The eigenenergies $E_{n}^{(2)}(R)$ ($n=1,2,3$) of the Hamiltonian
(\ref{eq:H_sp}) give the modified level configuration (e.g. the eigenstates
$|\phi_{n}^{(2)}\rangle$) for the Rydberg pair states, which strongly
depend on the coupling strength and detuning between the MW field
and the Rydberg states. As a specific example, we show, in Fig. \ref{fig:LevelConfig=00003D000026IP}(a),
the distance-dependent interactions between the Rydberg states $|58p\rangle$
and $|50s\rangle$ for Rb atoms. The Rydberg molecular potential appears at the upper (black solid) branch corresponding to the
pair state $|50s\rangle+|50s\rangle$ at large distances. It shows an
attractive character when the interatomic separation $R$ is larger
than 2 $\mu\text{m}$ and is repulsive
 at short distances. 
 

The ground state atoms are coupled from the ground state
$|5s\rangle$ to the MW coupled Rydberg states $|E_{+}^{(1)}\rangle=\text{sin}\left(\phi/2\right)|n^{\prime}s\rangle+\text{\text{cos}\ensuremath{\left(\phi/2\right)}}|np\rangle$
($|E_{-}^{(1)}\rangle=-\text{sin}\left(\phi/2\right)|np\rangle+\text{cos}\left(\phi/2\right)|n^{\prime}s\rangle$)
with the laser detunings $\Delta_{+}$($\Delta_{-}$) and coupling
strengths $\Omega_{+}=$$\text{ cos}\left(\phi/2\right)|\vec{d}_{5n}\cdot\vec{E}_{0}|/\hbar$
($\Omega_{-}=$$\text{ sin}\left(\phi/2\right)|\vec{d}_{5n}\cdot\vec{E}_{0}|/\hbar$),
where $\phi=\text{tan}^{-1}(2\Omega_{mw}/\Delta_{mw})$, $\vec{d}_{5n}$
is the dipole moment of the transition $|5s\rangle\leftrightarrow|np\rangle$,
and $\vec{E}_{0}$ is the electric field amplitude. We are interested
in the RMD based on two-photon dispersive interactions, that is, the
doubly-excitation states are far-off resonant from the two-photon
transition intermediated by $|E_{+}^{(1)}\rangle$. At large distances,
the effective Rabi frequencies for the transition channel $|5s\rangle|5s\rangle\rightarrow|5s\rangle|E_{+}^{(1)}\rangle\rightarrow|\phi_{1}^{(2)}\rangle$
($\sim|E_{+}^{(1)}\rangle|E_{+}^{(1)}\rangle$) is approximately given
by $\Omega_{2}\approx\Omega_{+}^{2}/\Delta_{+}$. While at short distances,
$\Omega_{2}(R)$ becomes distance-dependent. The parameter regime
for RMD is thus given by $|\Delta_{+}|\gg\Omega_{+}$ and $|2\Delta_{+}+E_{1}^{(2)}(\infty)-E_{1}^{(2)}(R)|\gg\Omega_{2},\gamma$,
from which we can then find the dressed interaction potential  via the perturbation theory (see the following section). The perturbation result of the dressed interaction
potential agrees with the full numerical calculation, shown in Fig. \ref{fig:LevelConfig=00003D000026IP}(b).

\section{Derivation of the effective master equation}
We consider two hyperfine ground states $|0\rangle$ and $|1\rangle$ where state $|1\rangle$ is weakly coupled to the Rydberg state
$|r\rangle=|+\rangle$ by the dressing laser (Rabi frequency $\Omega=\Omega_+$ and detuning
$\Delta=\Delta_+$). The Hamiltonian of the atoms is given by $H_{o}=H_{a}+H_{i}$,
where the atomic Hamiltonian for the spin states $|0\rangle$, $|1\rangle$
and Rydberg state $|r\rangle$ is 
\begin{equation}
H_{a}=\sum_{j}\left[\Delta\sigma_{rr}^{(j)}+\Delta_{0}\sigma_{00}^{(j)}+\left(\frac{g}{2}\sigma_{01}^{(j)}+\frac{\Omega}{2}\sigma_{1r}^{(j)}+\text{H.c.}\right)\right]
\end{equation}
where $\Delta_{0}$ and $g$ is the detuning and Rabi frequency between
the two hyperfine states (e.g. coupled through a MW field or a Raman process). The two-body interaction in the Rydberg state is 
\begin{equation}
H_{i}=\frac{1}{2}\sum_{jk}U(R_{jk})\sigma_{rr}^{(j)}\sigma_{rr}^{(k)},
\end{equation} 
where $U(R)=E_2^{(2)}$ is the molecular interaction potential of the MW coupled Rydberg state. Taking into account of the spontaneous decay
in the Rydberg state, the dynamics is governed by a master equation
\begin{equation}
\dot{\rho}=-[H_{o},\rho]-\gamma\sum_{j}\left(\sigma_{1r}^{(j)}\rho\sigma_{r1}^{(j)}-\frac{1}{2}\left\{ \sigma_{r1}^{(j)}\sigma_{1r}^{(j)},\rho\right\} \right).\label{eq:H3level}
\end{equation}

We first derive the effective master equation of two atoms. This is done in a regime where the detuning of the dressing laser is large, i.e. $\Delta>\Omega\gg\gamma\gg g\approx\Delta_{0}$.  We will then eliminate dynamics in the Rydberg state to obtain the effective dynamics by separating different
time scales. As the dynamics due to the dressing
laser is fast, this allows us to split the Hilbert space
into two parts, $P$ and $Q$ where $P$- and $Q$-subspace define
the slowly and rapidly varying quantities with $P+Q=\mathbb{1}$. The effective
Hamiltonian for the slow dynamics can be obtained through 
\begin{equation}
H_{eff}=PHP-PHQ\frac{1}{QHQ}QHP.
\end{equation}
In the two atom basis $\{|00\rangle,\,|01\rangle,\,|10\rangle,\,|11\rangle,\,|rr\rangle\}$,
the matrix form of the Hamiltonian is expressed as 
\begin{eqnarray*}
H_{eff}=\begin{pmatrix}2\Delta_{0} & 0 & 0 & 0 & 0\\
0 & \Delta_{0}-\frac{\Omega^{2}}{4\Delta} & 0 & 0 & 0\\
0 & 0 & \Delta_{0}-\frac{\Omega^{2}}{4\Delta} & 0 & 0\\
0 & 0 & 0 & -\frac{\Omega^{2}}{2\Delta} & -\frac{\Omega^{2}}{2\Delta}\\
0 & 0 & 0 & -\frac{\Omega^{2}}{2\Delta} & U+2\Delta-\frac{\Omega^{2}}{2\Delta}
\end{pmatrix}
\end{eqnarray*}

To study the dissipative process, we assume decay of Rydberg states
is fast than the coherent couplings. To consider different time scales,
the master equation $\dot{\hat{\rho}}=(\mathcal{L}_{0}+\mathcal{L}_{1})\hat{\rho}$
is split into the fast (denoted by $\mathcal{L}_{0}\hat{\rho}$) and
slow (denoted by $\mathcal{L}_{1}\hat{\rho}$) parts, where 
\begin{eqnarray}
\frac{\mathcal{L}_{0}\hat{\rho}}{\gamma} & = & \sum_{j=1,2}\left(\hat{\sigma}_{1r}^{j}\hat{\rho}\hat{\sigma}_{r1}^{j}-\frac{1}{2}\{\hat{\rho},\hat{\sigma}_{r1}^{j}\hat{\sigma}_{1r}^{j}\}\right),\nonumber \\
\mathcal{L}_{1}\hat{\rho} & = & -i[\hat{H}_{eff},\hat{\rho}].
\end{eqnarray}
We will trace the fast dynamics and derive an effective master equation
for the slow dynamics via the second order perturbation calculation.

Here we define a projection operator $\mathcal{P}_{0}=\lim_{t\to\infty}e^{t\mathcal{L}_{0}}$,
which projects the density matrix to the subspace corresponding to
the relatively slow dynamics, i.e. $\hat{\rho}_{0}=\mathcal{P}_{0}\hat{\rho}$.
The first order correction to the slowly varying dynamics vanishes
$\mathcal{P}_{0}\mathcal{L}_{1}\mathcal{P}_{0}\hat{\rho}=0$. We then
calculate the second order correction $-\mathcal{P}_{0}\mathcal{L}_{1}(\mathcal{I}-\mathcal{P}_{0})\mathcal{L}_{1}\mathcal{P}_{0}\hat{\rho}$.
A tedious but straightforward calculation yields an effective master
equation depending on the two-atom dephasing, 
\begin{equation}
\dot{\hat{\rho}}_{e}\approx-i[\widetilde{V}_{12},\rho_{e}]+\text{\ensuremath{\gamma}}_{12}\left(\hat{\sigma}_{11}^{1}\hat{\sigma}_{11}^{2}\hat{\rho}_{e}\hat{\sigma}_{11}^{2}\hat{\sigma}_{11}^{1}-\frac{1}{2}\{\hat{\sigma}_{11}^{2}\hat{\sigma}_{11}^{1},\hat{\rho}_{e}\}\right).\label{eq:H2Eff}
\end{equation}
where $\gamma_{12}=\Omega^{4}\gamma/2\Delta^{2}[(U(R_{12})+2\Delta)^{2}+\gamma^{2}]$
and the dressed interaction $V_{12}=\Omega^{4}[U(R_{12})+2\Delta]/2\Delta^{2}[(U(R_{12})+2\Delta)^{2}+\gamma^{2}]$
without including the correction induced by the single-photon Stark
shifts.

\begin{figure}
\includegraphics[width=1\columnwidth]{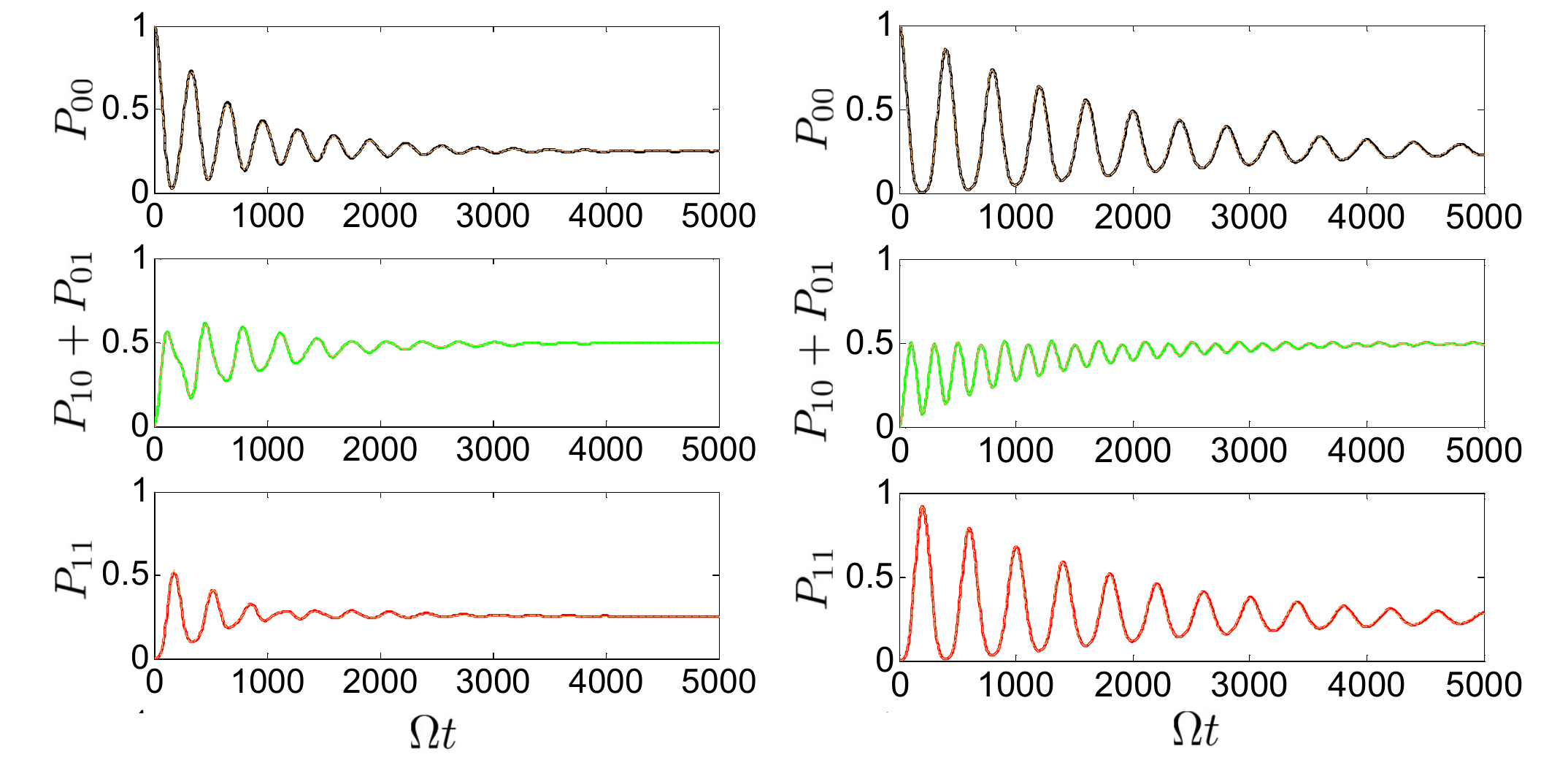}
\caption{\label{fig:Eff_Dyn_Dephasing} Dynamics of two interacting
atoms under RMD. Population of states $|00\rangle$, $\left(|01\rangle+|10\rangle\right)/\sqrt{2}$,
and $|11\rangle$ with the full three-level Hamiltonian (\ref{eq:H3level})
(solid) and the effective two-level model (\ref{eq:H2Eff}) (dash)
for $U(R_{12})/\Omega=21$ (left panel) and $U(R_{12})/\Omega=40$
(right panel), respectively. Note that the difference is barely seen in the numerical simulation. Other parameters are $\Omega=1$, $\Delta/\Omega=10$,
$\gamma=0.1$, $g=2.5\times10^{-3}$, and $\Delta_{0}=\Omega^{2}/4\Delta$.}
\end{figure}

To illustrate the accuracy of the perturbative calculation, we consider the system
dynamics under the Rydberg molecule dressing with $(U(R_{12}),\Delta)/\Omega=(21,10)$
(left panel) and $(U(R_{12}),\Delta)/\Omega=(40,10)$ (right panel),
respectively, shown in Fig. \ref{fig:Eff_Dyn_Dephasing}. The former leads to the dressed interaction $V_{12}/\Omega\approx2.5\times10^{-3}$
and the two-body decay rate $\gamma_{12}\sim\Omega^{2}\gamma/2\Delta^{2}=5\times10^{-3}\gamma$,
which is comparable to the effective single-photon decay $\gamma_{1}\sim\Omega^{2}\gamma/4\Delta^{2}$;
while for the latter, the dressed energy and the effective decay rate
are $V_{12}/\Omega\approx2.5\times10^{-4}$ and $\gamma_{12}\sim10^{-3}\gamma_{1}$,
respectively. The dissipative dynamics of the two-atom system is then
determined by the ratios with respect to the driving strength: $g/V_{12}$
and $g/(2\gamma_{1}+\gamma_{12})$. For the limit $V_{12}/g\gg1$,
the system can only transit between the states $|00\rangle$ and $|10\rangle$
($|01\rangle$), corresponding to the two-photon blockade, and the
two-body decay accelerates the convergence to the steady state. In
contrast, for $g/V_{12}\gg1$, the two-body decay effect can be largely neglected.

\section{Spin squeezing of the Rydberg dressed atoms}

Consider the effective decay described by the non-Hermitian (conditional)
Hamiltonian
\begin{eqnarray}
H_{c} & = & \sum_{i=1}^{N}\left[\frac{1}{2}\sum_{j\neq i}\overline{\kappa}_{ij}-i\frac{\gamma^{(1)}}{2}\right]J_{z}^{(i)}+\sum_{i<j}\overline{\kappa}_{ij}J_{z}^{(i)}J_{z}^{(j)}-i\frac{\Gamma_{G}}{2},
\end{eqnarray}
\begin{equation}
H_{m}=\sum_{i=1}^{N}\left(-\frac{\Delta_{m}}{2}J_{z}^{(i)}+\Omega_{m}J_{x}^{(i)}\right)+\text{H.c.},
\end{equation}
where
\[
\overline{\kappa}_{ij}=V(R_{ij})-i\frac{\gamma_{ij}^{(2)}}{2},
\]
\[
\Gamma_{G}=\frac{N\gamma^{(1)}}{2}+\frac{1}{2}\sum_{i<j}\frac{\gamma_{ij}^{(2)}}{2},
\]
with 
\[
\gamma_{ij}^{(2)}=\frac{\Omega^{4}\gamma}{2\Delta^{2}[(U(R_{ij})+2\Delta)^{2}+\gamma^{2}]}
\]
being the distance-dependent two-atom collective decay rate.

Under the mean-field approximation, the collective decay can be rewritten
in the following form,
\begin{align}
\sum_{j<k}\gamma_{jk}^{(2)}\hat{\sigma}_{n}^{(j)}\hat{\sigma}_{n}^{(k)}\approx & \frac{1}{2}\sum_{j\neq k}\gamma_{jk}^{(2)}\left(\hat{\sigma}_{n}^{(j)}\langle\hat{\sigma}_{n}^{(k)}\rangle+\langle\hat{\sigma}_{n}^{(j)}\rangle\hat{\sigma}_{n}^{(k)}-\langle\hat{\sigma}_{n}^{(j)}\rangle\langle\hat{\sigma}_{n}^{(k)}\rangle\right)\nonumber \\
= & \frac{1}{2}\sum_{j}\hat{\sigma}_{n}^{(j)}\left(\sum_{k\neq j}\gamma_{jk}^{(2)}\langle\hat{\sigma}_{n}^{(k)}\rangle\right)+\frac{1}{2}\sum_{j}\langle\hat{\sigma}_{n}^{(j)}\rangle\left(\sum_{k\neq j}\gamma_{jk}^{(2)}\hat{\sigma}_{n}^{(k)}\right)\nonumber \\
 & -\frac{1}{2}\sum_{j}\langle\hat{\sigma}_{n}^{(j)}\rangle\left(\sum_{k\neq j}\gamma_{jk}^{(2)}\langle\hat{\sigma}_{n}^{(k)}\rangle\right)\nonumber \\
= & \hat{\Gamma}(0)\langle\hat{\sigma}_{n}\rangle\sum_{j}^{N}\hat{\sigma}_{n}^{(j)}-\frac{N}{2}\hat{\Gamma}(0)\langle\hat{\sigma}_{n}\rangle^{2},
\end{align}
where $\hat{\Gamma}(\boldsymbol{q})$ is the Fourier transform of
the dissipation-rate matrix $\gamma_{jk}^{(2)}=\gamma^{(2)}(|\boldsymbol{R}_{j}-\boldsymbol{R}_{k}|)$:
\begin{equation}
\hat{\Gamma}(\boldsymbol{q})=\sum_{\boldsymbol{R}}\gamma^{(2)}(\boldsymbol{R})e^{-i\boldsymbol{q}\cdot\boldsymbol{R}}
\end{equation}
and therefore $\hat{\Gamma}(0)=\sum_{|k-j|=1}^{N-1}\gamma_{jk}^{(2)}$,
which is denoted as $\Gamma_{0}$ later for conciseness. For nearest
neighbor interactions only, one has $\Gamma_{0}=Z\gamma_{12}^{(2)}$,
where $Z$ is the lattice coordination number, i.e. the number of
nearest neighbors of any given site. For the one dimensional case,
we thus have
\begin{eqnarray}
H_{c} & = & \sum_{i=1}^{N}\left[-i\frac{\gamma^{(1)}}{2}+\frac{1}{2}\sum_{j\neq i}\left(V_{ij}-i\frac{\gamma_{ij}^{(2)}}{2}\right)\right]J_{z}^{(i)}+\sum_{i<j}\left(V_{ij}-i\frac{\gamma_{ij}^{(2)}}{2}\right)J_{z}^{(i)}J_{z}^{(j)}-i\frac{\Gamma_{G}}{2},\nonumber \\
 & = & \sum_{i=1}^{N}\left(\frac{1}{2}\sum_{j\neq i}V_{ij}\right)J_{z}^{(i)}+\sum_{i<j}V_{ij}J_{z}^{(i)}J_{z}^{(j)}-i\left[\frac{\gamma^{(1)}}{2}+\frac{\Gamma_{0}}{4}\right]\sum_{i=1}^{N}J_{z}^{(i)}\nonumber \\
 &  & -\frac{i}{2}\left[\Gamma_{0}\langle J_{z}^{(i)}\rangle\sum_{j}^{N}J_{z}^{(j)}-\frac{N}{2}\Gamma_{0}\langle J_{z}^{(i)}\rangle^{2}\right]-i\frac{\Gamma_{G}}{2}\nonumber \\
 & = & \sum_{j=1}^{N}\left(\frac{1}{2}\sum_{k\neq j}V_{jk}\right)J_{z}^{(j)}+\sum_{j<k}V_{jk}J_{z}^{(j)}J_{z}^{(k)}-i\frac{N}{4}\left[\gamma^{(1)}+\Gamma_{0}\left(\frac{1}{4}-\bar{\langle J_{z}\rangle}{}^{2}\right)\right]\nonumber \\
 &  & -\frac{i}{2}\left[\gamma^{(1)}+\Gamma_{0}\left(\frac{1}{2}+\bar{\langle J_{z}\rangle}\right)\right]\sum_{j=1}^{N}J_{z}^{(j)},\nonumber \\
 & = & \frac{1}{2}\sum_{j=1}^{N}\left(\sum_{k\neq j}V_{jk}-i\Gamma_{z}\right)J_{z}^{(j)}+\sum_{j<k}V_{jk}J_{z}^{(j)}J_{z}^{(k)}-i\frac{\text{\ensuremath{\bar{\Gamma}}}}{2},\label{eq:MF_H}
\end{eqnarray}
where
\[
\text{\ensuremath{\bar{\Gamma}}}=\frac{N}{2}\left[\gamma^{(1)}+\Gamma_{0}\left(\frac{1}{4}-\bar{\langle J_{z}\rangle}{}^{2}\right)\right],
\]
\[
\Gamma_{z}=\gamma^{(1)}+\Gamma_{0}\left(\frac{1}{2}+\bar{\langle J_{z}\rangle}\right),
\]
and $\bar{\langle J_{z}\rangle}$ is the mean value of the $z$-component
of the angular momentum for the individual atom. For numeric,
we evaluate $\bar{\langle J_{z}\rangle}$ by the weighted mean for
all spins and the time average for the implementation period $\tau$,
namely,
\[
\overline{\langle J_{z}(\tau)\rangle}=\frac{1}{N\tau}\int_{0}^{\tau}d\tau\langle J_{z}(\tau)\rangle,
\]
which will be analytically calculated in the Heisenberg picture and
should be effective in the mean-field regime.

We measure the degree of spin squeezing by the parameter proposed
by Wineland et al. \cite{Wineland1994},

\begin{equation}
\xi^{2}=\frac{N\left(\Delta J_{\vec{n}_{\bot}}\right)^{2}}{|\langle\vec{J}\rangle|^{2}},\label{eq:Squeezing_Wineland}
\end{equation}
where 
\begin{equation}
\langle\vec{J}\rangle=\langle\sum_{k=1}^{N}J^{(k)}\rangle=\sqrt{\langle J_{x}\rangle^{2}+\langle J_{y}\rangle^{2}+\langle J_{z}\rangle^{2}}\hat{\varepsilon}_{z}
\end{equation}
is the mean collective angular momentum of all spins with its direction
$\hat{\varepsilon}_{z}$ being predefined along the $z$-axis, corresponding
to the initial unitary population of the ground state for all atoms,
and 
\begin{equation}
J_{\vec{n}_{\bot}}\left(\theta\right)=\textrm{cos}\left(\theta\right)J_{x}+\textrm{sin}\left(\theta\right)J_{y}
\end{equation}
is the perpendicular spin component with $\theta$ being the angle
with respect to the $x$ axis. Thus, the fluctuations in $J_{\vec{n}_{\bot}}\left(\theta\right)$
can be calculated by
\begin{align}
\left(\Delta J_{\vec{n}_{\bot}}\right)^{2}= & \langle J_{\vec{n}_{\bot}}^{2}\left(\theta\right)\rangle-\langle J_{\vec{n}_{\bot}}\left(\theta\right)\rangle^{2}\nonumber \\
= & \left(\textrm{cos}^{2}\left(\theta\right)\langle J_{x}^{2}\rangle+\textrm{sin}^{2}\left(\theta\right)\langle J_{y}^{2}\rangle+\text{sin}\left(\theta\right)\textrm{cos}\left(\theta\right)\langle J_{x}J_{y}+J_{y}J_{x}\rangle\right)\nonumber \\
 & -\left(\textrm{cos}^{2}\left(\theta\right)\langle J_{x}\rangle^{2}+\textrm{sin}\left(\theta\right)\langle J_{y}\rangle^{2}+2\text{sin}\left(\theta\right)\textrm{cos}\left(\theta\right)\langle J_{x}\rangle\langle J_{y}\rangle\right),\nonumber \\
= & \textrm{cos}^{2}\left(\theta\right)\left(\Delta J_{x}\right)^{2}+\textrm{sin}^{2}\left(\theta\right)\left(\Delta J_{y}\right)^{2}\nonumber \\
 & +\text{sin}\left(\theta\right)\textrm{cos}\left(\theta\right)\left(\langle J_{x}J_{y}+J_{y}J_{x}\rangle-2\langle J_{x}\rangle\langle J_{y}\rangle\right).
\end{align}
Without considering the effect of the atomic spontaneous emission
(i.e., $\gamma^{(1)}=0$ and $\gamma_{ij}^{(2)}=0$), and based on
the spin-echo protocol, one can find mean values of the spin components analytically~\cite{Gil2014}:
\begin{equation}
\langle J_{x}\rangle=\langle J_{y}\rangle=0,\text{ }\langle J_{z}\rangle=\langle\sum_{i=1}^{N}J_{z}^{(i)}\rangle=-\frac{1}{2}\sum_{i}^{N}\prod_{j\neq i}^{N}\textrm{cos}(\varphi_{ij}),\text{ }
\end{equation}
\begin{align}
\langle J_{x}^{2}\rangle= & \langle\sum_{k=1}^{N}J_{x}^{(k)}\cdot\sum_{l=1}^{N}J_{x}^{(l)}\rangle=\frac{N}{4}+2\sum_{k<l}^{N}\langle J_{x}^{(k)}J_{x}^{(l)}\rangle\nonumber \\
= & \frac{N}{4}+\frac{1}{4}\sum_{i<j}^{N}\left[\prod_{k\neq i,j}^{N}\textrm{cos}(\varphi_{ijk}^{-})-\prod_{k\neq i,j}^{N}\textrm{cos}(\varphi_{ijk}^{+})\right],
\end{align}

\begin{equation}
\langle J_{y}^{2}\rangle=\frac{N}{4}+\underbrace{2\sum_{i<j}^{N}\langle J_{y}^{(i)}\cdot J_{y}^{(j)}\rangle}_{0}=\frac{N}{4},
\end{equation}
and
\begin{equation}
\langle J_{x}J_{y}+J_{y}J_{x}\rangle=-\sum_{i<j}^{N}\textrm{sin}(\varphi_{ij})\prod_{k\neq i,j}^{N}\textrm{cos}(\varphi_{ik}),
\end{equation}
where $\varphi_{ij}=V_{ij}\tau/2$ and $\varphi_{ijk}^{\pm}=\varphi_{ik}\pm\varphi_{jk}.$

When we take into account the non-Hermitian Hamiltonian (\ref{eq:MF_H}),
where the decoherence is estimated by the mean-field single-body
decay, and redo the spin-echo protocol by using the following transformations

\begin{align}
e^{i\alpha^{*}J_{z}^{(k)}}J_{x}^{(k)}e^{-i\alpha J_{z}^{(k)}}= & \text{cos}(\text{Re}\alpha)J_{x}^{(k)}-\text{\text{sin}(\text{Re}\ensuremath{\alpha})}J_{y}^{(k)},\\
e^{i\alpha^{*}J_{z}^{(k)}}J_{y}^{(k)}e^{-i\alpha J_{z}^{(k)}}= & \text{cos}(\text{Re}\alpha)J_{y}^{(k)}+\text{\text{sin}(\text{Re}\ensuremath{\alpha})}J_{x}^{(k)},\\
e^{i\alpha^{*}J_{z}^{(k)}}J_{z}^{(k)}e^{-i\alpha J_{z}^{(k)}}= & e^{2\text{Im}\alpha J_{z}^{(k)}}J_{z}^{(k)},
\end{align}
the mean values and the quantum fluctuations of the spin angular momentum
operators are alternatively given by
\begin{equation}
\langle J_{m}\rangle\rightarrow e^{-\text{\ensuremath{\bar{\Gamma}}}\tau}\langle J_{m}\rangle,\text{ }\langle J_{m}J_{n}\rangle\rightarrow e^{-\text{\ensuremath{\bar{\Gamma}}}\tau}\langle J_{m}J_{n}\rangle,\label{eq:AM_NH}
\end{equation}
and
\begin{align}
\left(\Delta J_{\vec{n}_{\bot}}\right)^{2}= & e^{-\text{\ensuremath{\bar{\Gamma}}}\tau}\left[\textrm{cos}^{2}\left(\theta\right)\langle J_{x}^{2}\rangle+\textrm{si\ensuremath{n^{2}}}\left(\theta\right)\langle J_{y}^{2}\rangle\right]\nonumber \\
 & +\frac{1}{2}\text{sin}\left(2\theta\right)e^{-\text{\ensuremath{\bar{\Gamma}}}\tau}\langle J_{x}J_{y}+J_{y}J_{x}\rangle,
\end{align}
where $m,\text{ }n=x,$ $y$, or $z$, $\tau$ is the total optical
dressing time, and we have assumed the rotation pulses is fast
enough, such that the dissipation is negligible during the operation.
Moreover, the dissipative effects in the $z$-component of the individual
spin angular momentum {[}$J_{z}^{(k)}(\tau)\sim e^{-\Gamma_{z}\tau}${]}
during the two halves of optical dressing time counteract each other
due to spin-flip induced by the rotation $\pi$ pulse. Using the above results, we then calculate the squeezing parameter for different number of sites and interactions (RMD and SRD).

\end{document}